# Heterophased grain boundary-rich superparamagnetic Iron Oxides/carbon composite for Cationic and Anionic Dye Removal

K Priyananda Singh[1], Boris Wareppam[1], Raghavendra K G[2], N. Joseph Singh[1], A. C. de Oliveira[3], V. K. Garg[3], Subrata Ghosh[4], L. Herojit Singh[1]

[1] *Department of Physics, National Institute of Technology Manipur, Langol-795004, India.*
[2] *Department of Physics, Manipal Institute of Technology, Manipal Academy of Higher Education, Manipal, Karnataka, India-576104*
[3] *Institute of Physics, University of Brasília, 70919-970 Brasília, DF, Brazil.*
[4] *Micro and Nanostructured Materials Laboratory – NanoLab, Department of Energy, Politecnico di Milano, via Ponzio 34/3, Milano - 20133, Italy*

**ABSTRACT**

Iron oxide-based nanostructures receive significant attention as an efficient adsorbent for organic dyes removal. The removal properties have strong dependency on the stoichiometry, phases, reactive edges, defect states etc present in the iron-oxides nanostructures. Herein, iron oxide/carbon composite with well-defined heterophased grain boundaries is synthesized by simple precipitation method and followed by calcination. The local structure, spin dynamics and magnetic properties of heterophased iron oxides/carbon composite are thoroughly investigated to explore its cationic and anionic dye removal capability. To validate the effectivity of the presence of heterogeneous grain boundaries, iron oxide/carbon nanocomposite with homogeneous grain boundaries is also examined. It was found that the hetero-phased iron oxide/carbon showed removal capacity of 35.45 mg g$^{-1}$ and 45.84 mg g$^{-1}$ for cationic (Crystal Violet) and anionic (Congo Red) dyes, respectively as compared to that of as-synthesised imidazole-capped superparamagnetic α-Fe$_2$O$_3$ (25.11 mg g$^{-1}$ and 40.44 mg g$^{-1}$, respectively) and homophased iron oxide/carbon nanocomposite (9.41 mg g$^{-1}$ and 5.43 mg g$^{-1}$, respectively). The plausible mechanism on the local structural evolution of the heterophase in the course of calcination and increase of the removal capacity is discussed. A detailed dye adsorption investigation is presented including the adsorption kinetic study. The pseudo-second order kinetic model is found to be an appropriate one and suggests that the chemisorption is dominant factor leading to adsorption of dyes. Whereas Weber-Morris model indicate the strong influence of boundary layers of nanocomposite on the adsorption process.

Corresponding author email address:
subrata.ghosh@polimi.it or subrata.ghoshk@rediffmail.com (S.G.).
herojit@nitmanipur.ac.in or loushambam@gmail.com, (L.H.S.)

# 1. Introduction

The large amount of organic dyes usage in different industries like textile, paper, printing, photography, leather, cosmetic and huge effluents in water bodies is one of the major causes of environmental pollution[1,2]. A trace amount of dye (less than 1 ppm) can have significant effect and persistent for long in water[3]. Owing to their high resistance, stability, toxicity and inability to degrade easily, organic dyes mixing with water bodies pose various problems such as reduction of light penetration, decrease in visibility, increase in chemical oxygen demand (COD), and affects the overall aquatic eco-system severely. These problems demand the urgent need of developing clean, efficient and environment friendly technique to remove the organic pollutants from water bodies[4].

In reality, most of the industrial effluents include both cationic dyes (Crystal Violet, Methylene Blue etc.) and anionic dyes (Congo red, methyl orange, Eriochrome black T etc.). Although metal oxide-based nanostructures with various morphology and exciting physico-chemical properties are anticipated as promising materials for removal of the cationic and anionic dyes [5–7], they are highly selective in either of dyes and rejected the other types of dyes [8,9]. Amongst, $\gamma$-$Fe_2O_3$ and $\alpha$-$Fe_2O_3$ nanoaparticles (NPs) are much popular owing to their unique features, such as great biocompatibility, low synthetic cost and simplicity of functionalization, large magnetic susceptibility, large specific surface area, chemical stability, biocompatibility, amphoteric surface function, excellent catalytic adsorption with greater regenerative capacity [10,11]. For example, flower-like core-shell $Fe_3O_4$@$MnO_2$ composites, synthesized by hydrothermal process, could exhibit removal percentage of 95% for Congo Red (CR), but its removal efficiencies were below 15% for Crystal Violet (CV), Methyl orange (MO), Methylene blue (MB) and Rhodamine B (RB)[12]. Moreover, $Fe_3O_4$ NPs were functionalized by co-precipitation method using multi-carboxylate organic ligands, which exhibits 97% removal in CR dyes but, only 14% removal for CV and 9% removal for MO dyes[13]. A Zero-valent iron loaded biopolymer-based composites exhibited excellent removal efficiency for both anionic (CR) dye and cationic (CV) dye in addition to the ease of separation for reusability and greater stability. However, iron nanoparticles (NPs) are prone to oxidation on being exposure to air [14]. The poor stability of NPs leads to aggregation of adsorbents and often limits their application in dye removal. Another major issue with these magnetic iron oxide NPs ($\gamma$-$Fe_2O_3$ and $Fe_3O_4$) that they are often transformed into non-magnetic iron oxide phases like hematite ($\alpha$-$Fe_2O_3$) under various environmental conditions over a long period of time or under the influence of high temperature[15].

To tackle this challenge, surface coating of these NPs with polymers, carbon layers and organic ligands are the probable solution in addition to enhancing the surface reactivity and adsorption capacity further [16–19]. For instance, Saiphaneendra *et.al.* investigated the synergistic effect of hematite and magnetite particles on GO for methylene blue adsorption. The strong bonding between the embedded NPs and GO and possible coexistence of hematite and magnetite were proposed to be beneficial for dye removal application[20]. The enhanced dye adsorption by carbon compounds has an inherent limitation of carbon, being non-magnetic and not suitable for separation for reusability. Interestingly, the presence of both magnetic and non-magnetic interlayer can also alter the interphase coupling and reversal mechanism, which plays a crucial role for synchronous behaviour of interphase materials [21,22]. It is reported that the presence of carbon content in the heterophase junction, can be tuned by adjusting the activation temperature of the precursor, which uplift the photocatalytic hydrogen evolution [23]. Composite with heterophase grain boundary or interphase structures in nanomaterials can affect the physicochemical properties of nanomaterials owing to their high energy nature and changing electronic environment along the inter phase boundaries [24]. Basically, engineering the heterophased grain boundaries in nanocomposites may lead to creation of lattice defects and modification of charge carriers along the boundaries. Thus, these charged states and lattice defects may provide favourable conditions for attacking positively and negatively charged dyes simultaneously and the heterophased grain boundaries serve as an active sites to bind the chemical groups mediating dye adsorption effectively [25,26].

With this background, some important questions that will be addressed in this article:

(i) Is it possible to design an iron oxide/carbon nanocomposite with heterophased grain boundaries (hetero-IOCC), which will be stable under toxic dye environment without transforming its phase?
(ii) What could be the effectiveness of hetero-IOCC to remove both cationic and anionic dyes compared to its homogeneous counterpart (IOCC with homogeneous grain boundaries) or functionalized IONPs?
(iii) How does the changes in the local structure, magnetic properties, and catalytic properties of hetero-IOCC and homo-IOCC effect the performances of dye removal?

Focusing to address these critical questions, carbon containing iron oxide with heterophased grain boundaries is synthesized using thermal treatment on imidazole capped superparamagnetic (SPM) $\alpha$–$Fe_2O_3$. The local structural evolution from imidazole capped SPM $\alpha$–$Fe_2O_3$, to hetero-IOCC and then homo-IOCC is investigated with the help of X-ray diffraction (XRD), high-resolution transmission electron microscopy (HRTEM), Fourier transform infrared spectrometer (FTIR), X-ray photoelectron spectroscopy (XPS), Vibrating Sample Magnetometer (VSM), Thermo-gravimetric analysis (TGA) and Mössbauer spectroscopy. The plausible mechanism of local structure evolution during the decomposition of the imidazole attached on the surface of $\alpha$–$Fe_2O_3$ is discussed. We envisaged the suitability of iron-oxide/carbon composite with heterophased grain boundaries towards its potential utilization as an active material to remove both cationic and anionic dye over the homo-IOCC and bare imidazole capped SPM $\alpha$–$Fe_2O_3$.

## 2. Experimental methods
### 2.1. Materials

$Fe(NO_3)_3.9H_2O$ (98% purity), purchased from Merck Life Science Pvt. Ltd., India, and 2 methyl imidazole (HMIM), purchased from Sigma Aldrich (99% purity), were used as the precursors. CV dye and CR dye were from HiMedia Laboratories Pvt. Ltd. All the reagents used in the experiment were employed without further purification and they were of analytical grade. Double Distilled water was used in whole experimental process.

### 2.2 Nanocomposite preparation

An aqueous solution of $Fe(NO_3)_3.9H_2O$ (4.2 M) was mixed dropwise to the aqueous solution of 2-HMIM of the same molarity. The two solutions were kept at magnetic stirring for 10 minutes until a dark reddish coloured precipitate was formed. The precipitate was subsequently separated and collected using centrifuge (6000 rpm for 5 mins). The collected precipitate was washed for several times to remove the excess unreacted imidazole. The imidazole capped iron oxide NPs so collected were dried in an oven for 6 hours at around 60 °C and named as Im-IONP. The as-synthesized Im-IONP samples were subjected to thermal treatment using a conventional oven at 300 °C for 1 hour duration with the heating rate of 5 °C/min. Finally, the samples were naturally cooled down at room temperature and powdered properly using a mortar pestle to get Hetero-IOCC (**Scheme 1**). A same procedure was carried out to obtain the Homo-IOCC except the calcination temperature of 400 °C.

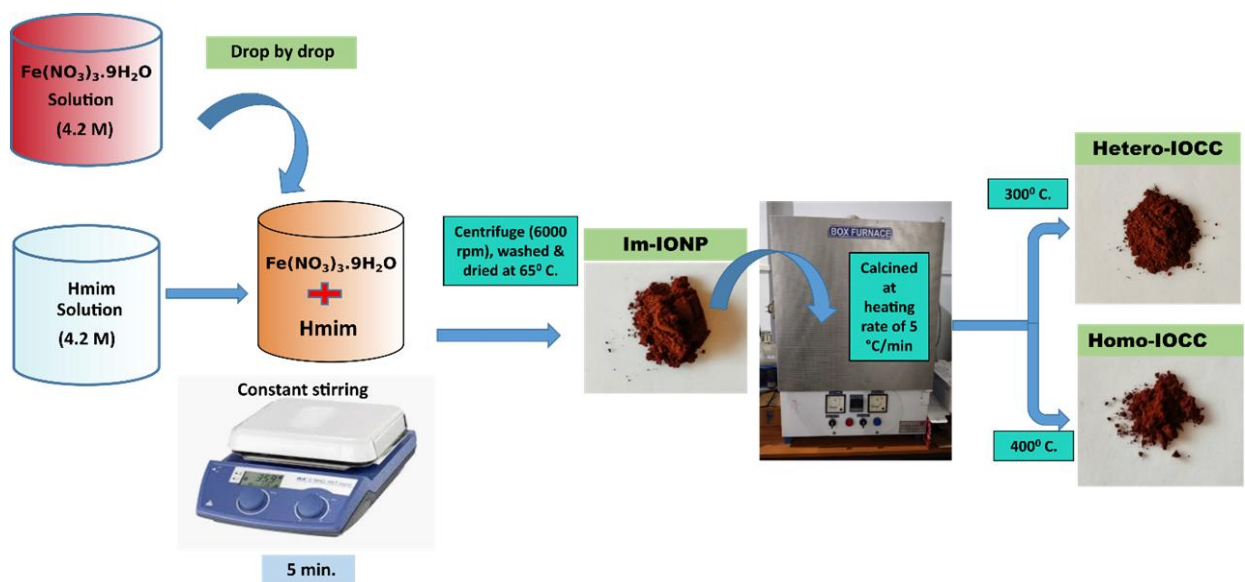

**Scheme 1**: *Schematics of preparation methodology of hetero-IOCC and homo-IOCC from Im-IONP*

*2.3 Material Characterization*

Phase and crystallographic information of the given sample were analyzed using XRD (Bruker-D8 advance-USA) with Cu K$_\alpha$ radiation of wavelength 1.54056 Å and HRTEM) (Tecnai G2, F30 with 300kV accelerating voltage). The particle size distribution analysis was carried out by measuring particle size using imageJ software. FTIR (Perkin Elmer-USA) was used to investigate the vibrational properties of the powders in attenuated total reflection (ATR) mode in the range 400-4000 cm$^{-1}$. Mössbauer spectroscopy was utilized in transmission mode by using a $^{57}$Co (Rh) source with initial activity of 25 mCi (Wissel, Germany). The magnetic characteristics of as-synthesised Im-IONP, Hetero-IOCC and Homo-IOCC were analyzed using Vibrating Sample Magnetometer (VSM-Model: LAKE SHORE VSM 7410-USA) at room temperature. XPS measurement was performed for sample on a PHI 5000 Versa Probe III- Japan equipped with Al Kα-945 monochromatic radiation source. TGA and Derivative thermogravimetry (DTG) were carried out for Im-IONP sample by a TGA (NETZSCH STA 449F3) equipped with a high-temperature furnace in the temperature range from 20 °C to 1000 °C at the heating rate of 10 °C/min (in argon gas environment).

*2.4 Adsorption experiments*

The adsorption study of CR and CV dyes was carried out by using UV-Vis spectrophotometer (Shimadzu-1800-Japan) in the wavelength range of 400 nm to 900 nm. The adsorption activities of the samples were assessed by measuring adsorption capacities and removal percentage (%) of CV and CR in an aqueous solution. For the batch adsorption experiment, 200 mg/L of Im-IONP, Hetero-IOCC and Homo-IOCC samples were added to initial dye concentration of 10 mg/L. The samples were kept in dark room without stirring and measurements were carried out to ensure that the equilibrium is achieved in dark room. The concentration of the dyes were calculated from the maximum absorbance peak corresponding to λ=498 nm for CR and λ=583 nm for CV. The experiment was done at different intervals of time for a duration of about 14 hours to attain the equilibrium. It was observed that a decrease in the absorbance was obtained within time.

The adsorption capacity was calculated based on the equation below:[27]

$$q_t = \frac{(C_0 - C_t)}{m} V \quad (1)$$

where $C_0$ and $C_t$ represent the initial concentration and the concentration at certain time interval of CV and CR respectively (mg/L), $V$ is the volume of the solution (mL) and m is the mass of the sample (mg). For equilibrium, $C_t = C_e$; $q_t = q_e$, where $C_e$ and $q_e$ are equilibrium concentration and equilibrium adsorption capacity.

The removal efficiency of various samples were calculated by using the relation:[28]

$$\text{Degradation Percentage (\%)} = \frac{C_0 - C_t}{C_0} \times 100 \quad (2)$$

## 3. Results and Discussion
*3.1 Structural analysis*

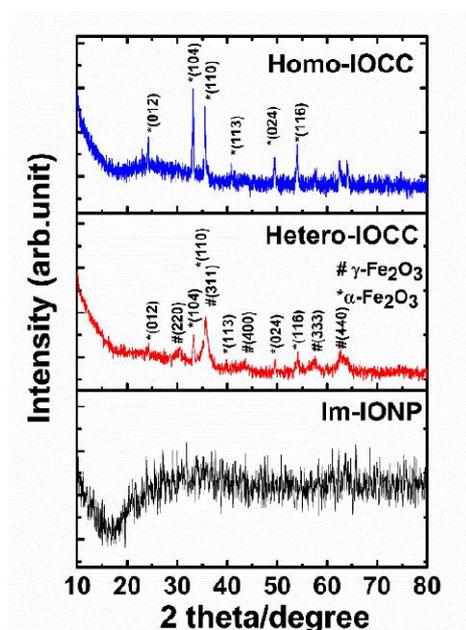

***Figure 1***: *XRD pattern of Hetero-IOCC along with the Im-IONP and Homo-IOCC.*

The powder XRD pattern of hetero-IOCC is shown in **Figure 1**. The observed characteristic diffraction peaks for hetero-IOCC at an angle 2θ = 24.2°, 33.2°, 35.6°, 40.8°, 49.4° are assigned to the scattering from (012), (104), (110), (113) and (024) planes of α-$Fe_2O_3$, respectively [JCPDS NO. 84-0306 (α-$Fe_2O_3$)][29,30]. Whereas, the respective diffraction peaks at an angle 2θ = 30.2°, 35.5°, 43.2° and 62.9° corresponding to the scattering from (220), (311), (400) and (440) planes of γ-$Fe_2O_3$ [JCPDS NO. 39-1346 (γ-$Fe_2O_3$)][29,30] were observed only in hetero-IOCC. On the other hand, homo-IOCC exhibits only the planes corresponding to the α-$Fe_2O_3$ (see Figure 1). The emergence of sharp peaks of higher intensities confirms the crystalline character of both hetero-IOCC and homo-IOCC. . The peaks corresponding to α-$Fe_2O_3$ components are sharp indicating a large crystallite size; on the other hand, the γ-$Fe_2O_3$ phases have comparatively broader peaks. Therefore, the α-$Fe_2O_3$ is likely to exhibit bulk-like magnetic properties, whereas γ-$Fe_2O_3$ is proposed to behave as a superparamagnet [31]. Hence, it is expected to have superparamagnet features for hetero-IOCC, whereas the bulk-like magnetic features for the homo-IOCC. This statement is further strengthened from the below VSM and Mössbauer studies. As can be seen from Figure 1, Im-IONP exhibits a scattered spectrum with no detectable diffraction peaks, indicating a lack of crystalline nature or a very fine SPM particle size [32].

Figure 2(a) shows the HRTEM micrographs of hetero-IOCC, which confirms the heterophased grain boundary of α-$Fe_2O_3$ and γ-$Fe_2O_3$ (represented by green and yellow lines). The radial profile analysis from selected area electron diffraction (SAED) pattern of hetero-IOCC (Figure 2b) shows that the rings correspond to α-$Fe_2O_3$ and γ-$Fe_2O_3$ phases. The corresponding (hkl) values (012), (104), (113), (024), (300) of α-$Fe_2O_3$ (represented by yellow colour) and (422), (311) of γ-$Fe_2O_3$ (represented by green colour) marked the presence of these two phases. Moreover, an inverse fast Fourier transform (IFFT) images corresponding to α-$Fe_2O_3$ and γ-$Fe_2O_3$ for hetero-IOCC samples are obtained through image processing and displays in Figure 2c. The heterophase grain boundaries of α-$Fe_2O_3$ and γ-$Fe_2O_3$ become less prominent in homo-IOCC (Figure 2d) as the (024), (300), (312) and (048) of α-$Fe_2O_3$ (represented by yellow ring) is dominated mostly (Figure 2e). In addition, the SAED pattern analysis of this sample shows the presence of the planes (000), (111), (5-3-1) and (6-20) of γ-$Fe_2O_3$ (represented by green colour) corresponding to γ-$Fe_2O_3$ as in Figure 2e. However, IFFT analysis of this sample (Figure 2f) shows the clear indication of the absence of γ-$Fe_2O_3$ phase in homo-IOCC sample.

The HRTEM micrographs of the Im-IONP indicates that the diffused lattice fringes from the fine nanocrystals can be observed at various locations (yellow circles in Figure 2g). Further, the SAED pattern shown in Figure 2h also shows diffused rings which confirms the nano-crystallinity of α-$Fe_2O_3$ in Im-IONP. This is in agreement with the XRD result showing scattered background and also resolve the limitation of XRD result where crystalline phases were not confirmed. The diffraction planes of these nano-crystallinity of α-$Fe_2O_3$ were identified to be (110), (202) and (300). IFFT image of the (113) oriented α-$Fe_2O_3$ crystals is shown in Figure 2(i) along with the *d*-spacing of 0.221 nm corresponding to it.

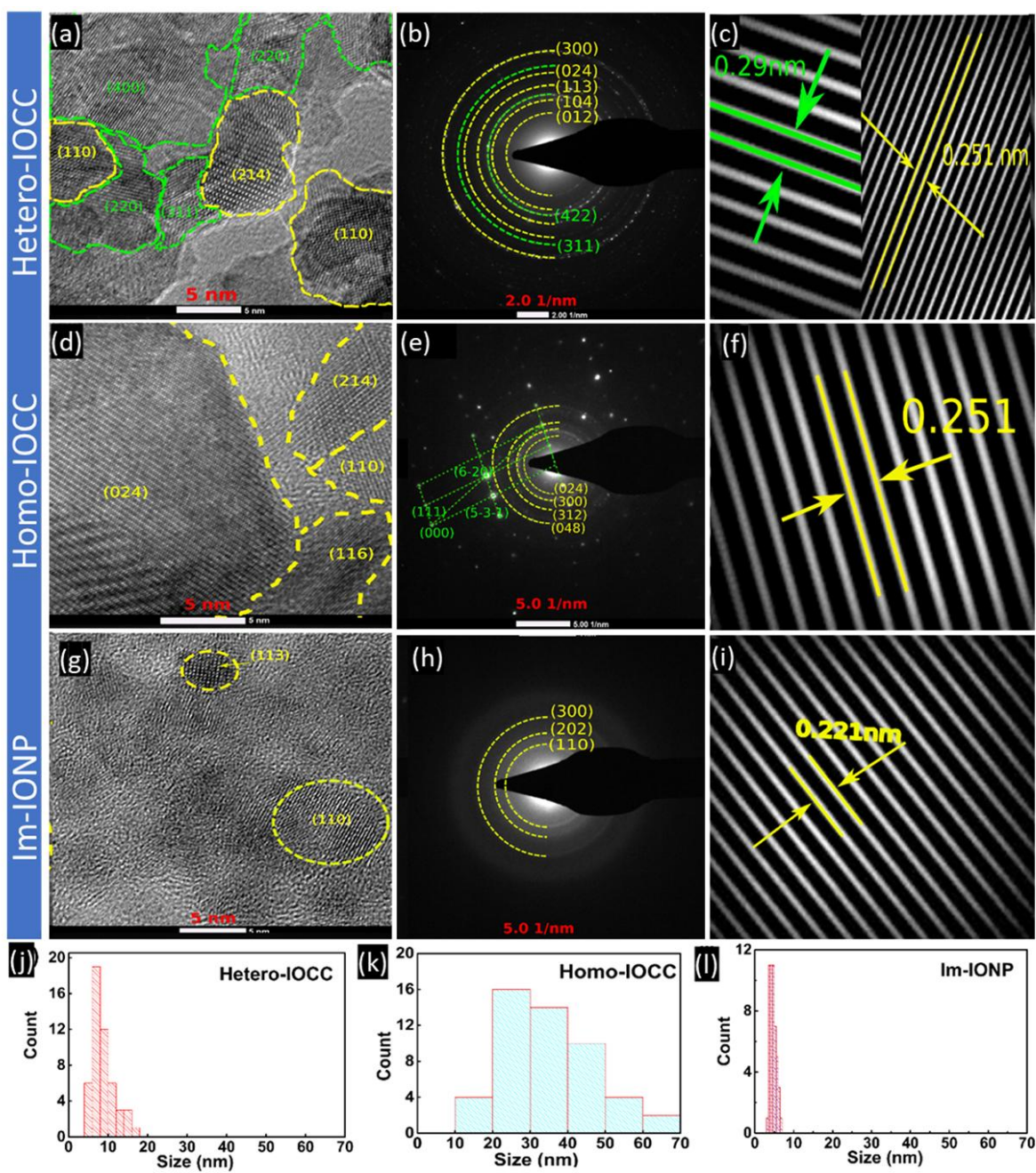

*Figure 2*: (a)HRTEM micrographs, (b) SAED patterns, and (c)IFFT of (110) plane of α-$Fe_2O_3$ and (220) plane of γ-$Fe_2O_3$ for hetero-IOCC; (d)HRTEM micrographs, (e) SAED patterns, and (f)IFFT of (110) plane of α-$Fe_2O_3$ for homo-IOCC; (g)HRTEM micrographs, (h) SAED patterns, and (i)IFFT of (113) plane of α-$Fe_2O_3$ for Im-IONP; The particle size distribution for (j) hetero-IOCC (k) homo-IOCC and (l) Im-IONP.

The particles size distributions of the samples were also carried out through image analysis of TEM micrographs of these three samples and the results are shown in Figure 2(j-l). As expected, the size of Im-IONP were ranging from 3 to 7 nm. Due to heat treatment, the resulting iron oxide NPs exhibit much greater particle sizes and the particles as clearly seen from Figure 2(g-i) indicate that the increment of particle size of samples follows the trend: homo IOCC (10-70 nm)

> hetero-IOCC (5-16 nm) > Im-IONP (3-7 nm), which can be attributed to the result of calcination at different temperature.

The observation of very fine particles is understood due to the instant capping of the precipitated α-$Fe_2O_3$ by 2-methyl imidazole thereby preventing them from agglomeration. The 2-methyl imidazole under thermal treatment releases reductive gases and reduces the α-$Fe_2O_3$ to cubic γ-$Fe_2O_3$ [33,34]. Thus, it can be said that the decomposition of imidazole from the surface of iron oxides NPs at 300 °C, nanocrystalline imidazole linked α-$Fe_2O_3$ NPs get reduced into γ-$Fe_2O_3$ NPs, developed a well-defined grain boundary and some of γ-$Fe_2O_3$ NPs, being exposed at the ambient atmosphere condition, get oxidized into α-$Fe_2O_3$ NPs. These all-simultaneous phenomenon results in a mixed γ-$Fe_2O_3$/ α-$Fe_2O_3$ phases with well-defined hetero-phased grain boundary (hetero-IOCC). These heterophased grain boundaries of γ-$Fe_2O_3$/ α-$Fe_2O_3$ become less prominent in homo-IOCC, which reveals that the heterophased structure get transformed mostly into single phase nanostructure of α-$Fe_2O_3$. Further quantification of this γ-$Fe_2O_3$ phase will be done using Mössbauer spectroscopy in the subsequent section of this article.

*3.2 Chemical Studies by XPS*

As the calcination was conducted at 300 °C and 400 °C, the presence of carbon on the surface of iron oxide is expected. Therefore, to understand the chemical composition and bonding states of different elements present in the material after decomposition, XPS was carried out on hetero-IOCC (Figure 3). The characteristic peaks of C, O, N and Fe at their corresponding binding energy values are clearly seen. The C1*s* spectrum comprises of the six peak positions with binding energies at around 283.7, 284.8, 286.1, 286.8, 288.1 and 289.1 eV corresponding to Fe-C, C-C, C-N, C-O, C=O and O-C=O, respectively[35,36]. Also, high resolution spectrum of Fe2*p* spectra indicates two spin orbit coupling signatures. The corresponding peaks at 710.8 eV and 713.3 eV represent the $Fe^{2+}$ 2$p_{3/2}$ and $Fe^{3+}$ 2$p_{3/2}$ orbits, whereas the characteristic peaks 724.3 eV and 726.2 eV arise due to the $Fe^{2+}$ 2$p_{1/2}$ and $Fe^{3+}$ 2$p_{1/2}$ orbits, respectively. This data is in agreement with the XRD and TEM result which predicted the structural transformation of α-$Fe_2O_3$ into γ-$Fe_2O_3$ in hetero-IOCC. Along with these four major peaks, there are two shake-up satellite peaks observed at 719.6 eV and 732.2 eV. Here, greater area of 2$p_{3/2}$ as compared to 2$p_{1/2}$ may be due to higher spin orbit *j-j* coupling having degeneracy 4 of Fe 2$p_{3/2}$ as compared to degeneracy 2 of Fe 2$p_{1/2}$. In the O1*s* spectra, the main characteristic peaks at 529.07 eV can be attributed to Fe-O bond. The binding energy peaks at 531.2 eV arises out of the oxygen defects in the iron oxide complex of hetero-IOCC sample[37]. The broadened peaks at 533.4 eV can be assigned to C-O bond [38]. XPS spectra of N1*s* is not so significant as compared to other spectra of C1*s*, O1*s* and Fe2*p*; however, the presence of nitrogen bond at binding energy range of around 395 eV to 402 eV can be detected.

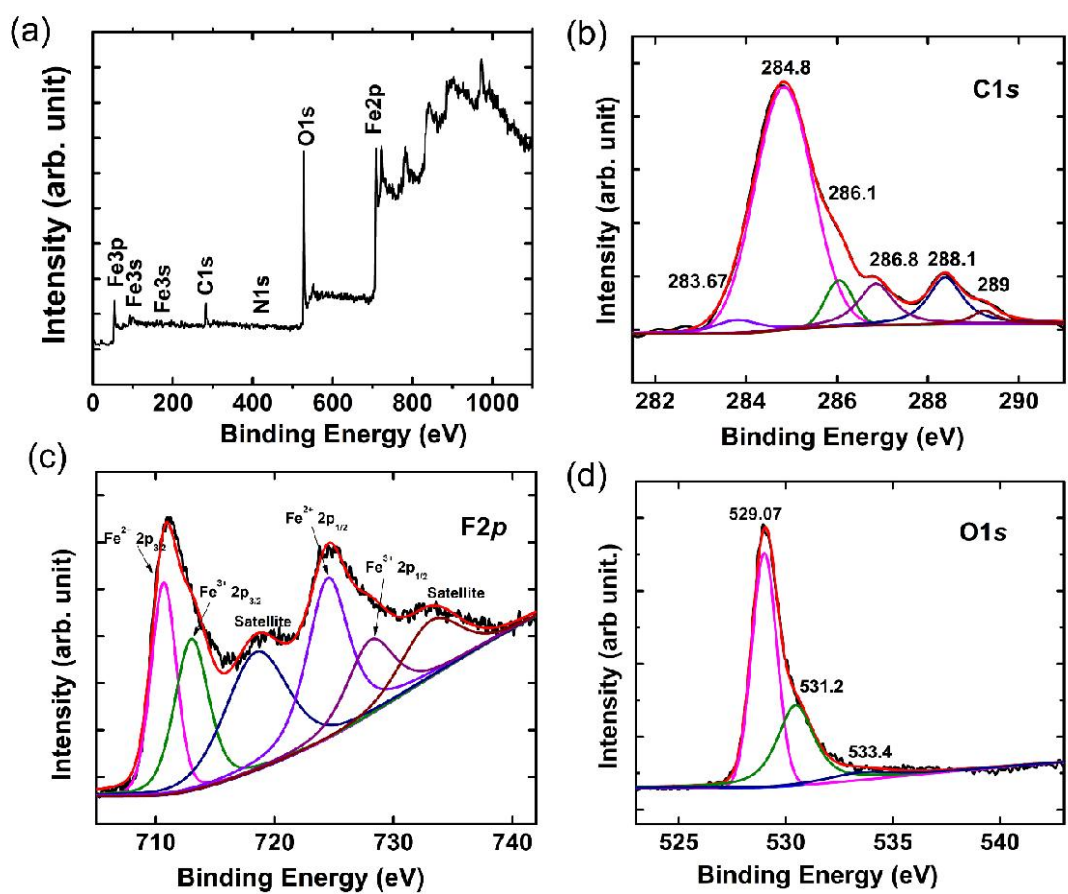

**Figure 3**: (a) Full spectra, (b) C 1*s* spectra, (c) Fe 2*p* spectra, (d) O 1*s* spectra of hetero-IOCC.

*3.3 Magnetic properties*

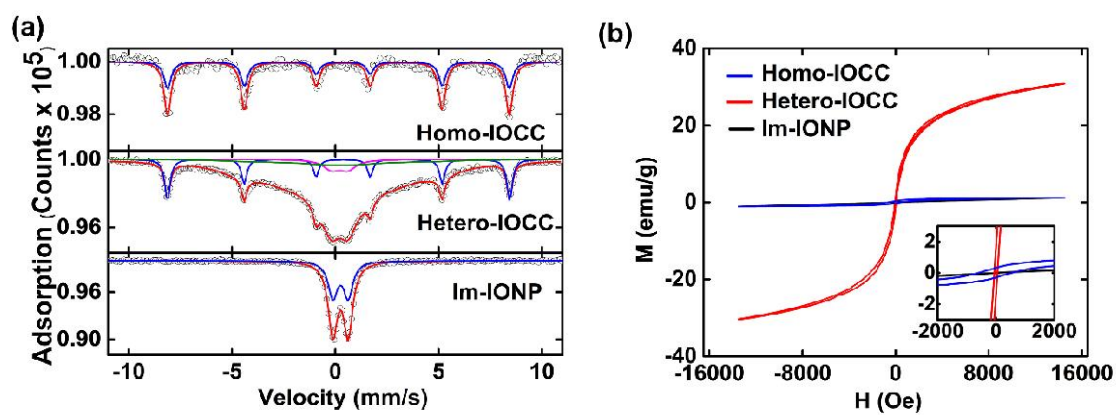

**Figure 4**: (a) Mössbauer spectra and (b) Magnetic hysteresis curve for hetero-IOCC along with Im-IONP and homo-IOCC.

**Table 1.** Room Temperature Mössbauer data of hetero-IOCC, homo-IOCC and Im-IONP. Data in bracket represents the standard deviation in decimal point.

| Sample | Area (%) | Isomer shift δ (mm/s) | Mean Hyperfine field, $B_{hf}$ (T) | Quadrupole splitting, Δ (mm/s) | Saturation Magnetization, $M_S$ (emu/g) | Coercivity, $H_C$ (Oe) |
|---|---|---|---|---|---|---|
| Hetero-IOCC | 13 | 0.21 (0) | | 0.79 (1) | 30.59 | 52.80 |
| | 13 | 0.26 (0) | 51.0 (1) | 0.22 (0) | | |
| | 74 | 0.23 (2) | 29.0 (3) | 0.01 (3) | | |
| Homo-IOCC | 100 | 0.26 (0) | 51.3 (2) | 0.23 (0) | 1.18 | 557.40 |
| Im-IONP | 100 | 0.25 (0) | | 0.73 (0) | 1.15 | 50.99 |

The Mössbauer spectra of Im-IONP consist of a SPM doublet experiencing **Δ** (Quadrupole splitting) and δ (isomer shift) of 0.73 mm/s and 0.25 mm/s, respectively (Table 1). In the Mössbauer studies, the doublet in Im-IONP transforms to two sextet and a doublet in hetero-IOCC (Figure 4a). The evolution of sextet with different $B_{hf}$ (Mean Hyperfine field) and the doublet in hetero-IOCC is due to varying distribution of particle size (5 – 16 nm) as a consequence of agglomeration during heat treatment and partial oxidation of γ-$Fe_2O_3$ to α-$Fe_2O_3$, which is also in agreement with TEM analysis. The SPM doublet of 13% is attributed to the isolated NPs with size less than 10 nm. The sextet with hyperfine field of 29 T consisting of 74 % of the total area. This fact is attributed from the particles exhibiting high interparticle magnetic interaction and those particles having dimensions larger than SPM regime but still experiencing large surface energy. This is plausible only when the HMIM surfactants, which prevent the particles from agglomeration, got decomposed resulting to the decrease in the interparticle distances and enhance the magnetic interaction[39]. The sextet with 13% of the total area experiencing at 51T, quadrupole splitting of 0.22 mm/s and isomer shift of 0.26 mm/s correspond to ideal bulk α-$Fe_2O_3$[40]. From the variation of hyperfine field, it is understood that the particles' agglomeration occurs during the decomposition of imidazole. In the course of oxidation to α-$Fe_2O_3$ from γ-$Fe_2O_3$, the rate of agglomeration or fusion of the oxidized particle is much higher than the nucleation of γ-$Fe_2O_3$, which are left unoxidized. The oxidized portion of the particles i.e. α-$Fe_2O_3$ from γ-$Fe_2O_3$ were often observed to have larger particle size as compared to the remaining metal oxides, as evident from TEM analysis. The agglomeration effect in oxidized α-$Fe_2O_3$ is further observed in the homo-IOCC, which was obtained by oxidising γ-$Fe_2O_3$ at 400 ºC. The Mössbauer spectrum consist of a single sextet exhibiting the properties of bulk α-$Fe_2O_3$ with no trace of SPM particles or γ-$Fe_2O_3$.

The hysteresis loop (Magnetisation 'M' *vs* Magnetising field 'H' curve) of hetero-IOCC confirms the behaviour of SPM ferromagnetism (Figure 4b). The increase in magnetisation may be due to the presence of ferrimagnetic γ-$Fe_2O_3$ phase. At lower magnetising field, it exhibits a ferromagnetic straight line with an abrupt increase in magnetisation. At higher magnetising field (14400 Oe), it exhibits unsaturated linear magnetisation, which is most probably due to the contribution of antiferromagnetic α-$Fe_2O_3$[41]. Hetero-IOCC experiences the magnetisation of 30.59 emu/g at the magnetising field of 14400 Oe. The magnetic transition from antiferro to ferromagnet is due to the structural transformation of α-$Fe_2O_3$ to ferrimagnetic γ-$Fe_2O_3$. The MH loop of Im-IONP, in Figure 4b, shows a linear and zero coercivity. From the Mössbauer spectrum as well MH loop, the Im-IONP is confirmed as antiferromagnetic α-$Fe_2O_3$ experiencing SPM. The MH loop of homo-IOCC shows a similar feature of Im-IONP. This indicates the anti-ferromagnetism of the particles i.e, α-$Fe_2O_3$. The coercivity ($H_c$) of the homo-IOCC is higher than the hetero-IOCC and Im-IONP, as indicated in Table 1 (inset of Figure 4b). Thus, Mössbauer spectrum of homo-IOCC consists of a single sextet corresponding to α-$Fe_2O_3$, as the presence of γ-$Fe_2O_3$ is below the detectable range from Mössbauer spectroscopy. It is observed that the magnetization of homo-IOCC drops significantly compared to hetero-IOCC, which may be due the oxidation of ferrimagnetic γ-$Fe_2O_3$ back to non-magnetic α-$Fe_2O_3$. This agrees with the Mössbauer data shown in Table 1.

*3.4 Hetero-IOCC formation mechanism*

To explore the formation mechanism of hetero-IOCC from Im-IONP, detailed FTIR studies were conducted. The Im-IONP precipitate was obtained as a consequence of the addition of 2-methyl imidazole, which acts as base, on the surface of NPs [42]. Figure 5a shows the FTIR spectra of Hmim, Im-IONP, Hetero-IOCC and Homo-IOCC samples. The peaks at 580 $cm^{-1}$ with a shoulder at around 626 $cm^{-1}$ corresponds to stretching vibration of Fe-O bond for Im-IONP. A comparative study of FTIR spectrums of Im-IONP and Hmm show similar vibration at 744 $cm^{-1}$, 826 $cm^{-1}$, 1110 $cm^{-1}$ and 1326 $cm^{-1}$ which correspond to N-H wagging, out-of-plane bending, C-N stretching vibrations and *in-plane* bending of imidazole ring, respectively. [43] This is understood as Hmim is attached to the surface of IONPs. However, a significant reduction in the peak intensity and a shift of imidazole peaks at 1440 $cm^{-1}$, which correspond to C-H rocking towards lower wavenumber in Im-IONP sample, were observed. This weakening of bond may be due to the interaction of metal ions with Hmim during the synthesis process [44]. Thus, the FTIR spectrums of hetero-IOCC and homo-IOCC show the absence of hydroxyl group (-OH) (~3440 $cm^{-1}$), C-H (methyl) (~2930 $cm^{-1}$) and C-H (IM) (~3137 $cm^{-1}$) vibrations. These may be attributed to the breakdown of imidazole ring, which can be further clarified from the TGA and DTG analysis. Two bands at 527 and 436 $cm^{-1}$ (Figure 5a) were emerged in IOCC, which can be attributed to the Fe–O vibration of γ-$Fe_2O_3$ and α-$Fe_2O_3$ [45,46].

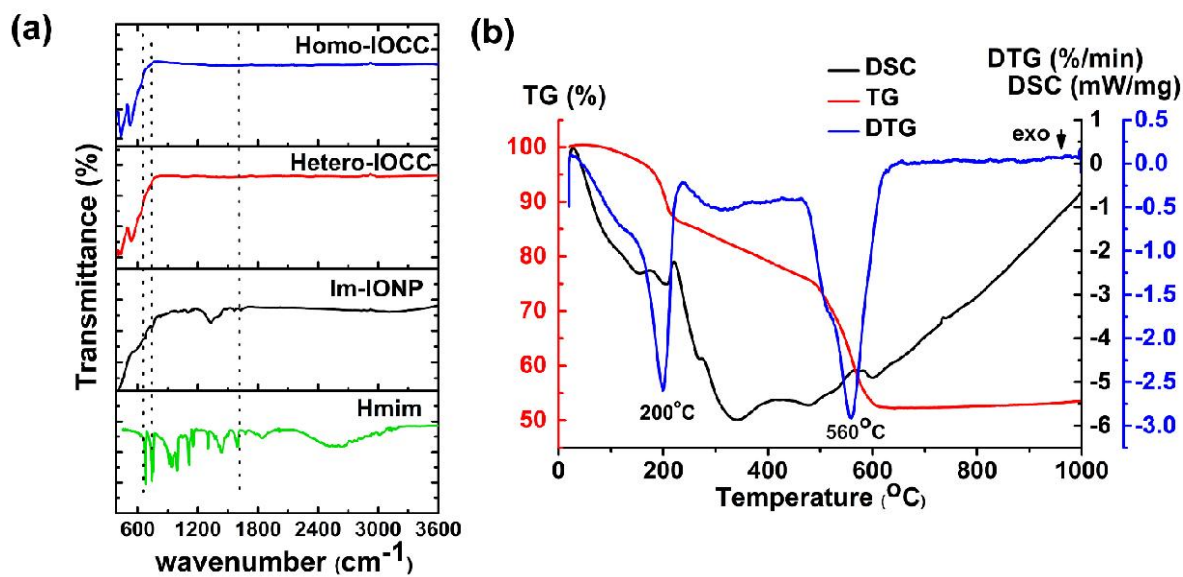

**Figure 5:** (a) FTIR of Hmim, Im-IONP, Hetero-IOCC and Homo-IOCC (b) TGA, DTG and DSC of Im-IONP.

To complement the decomposition of 2-methyl imidazole (Hmim) and reduction of α-$Fe_2O_3$, the imidazole attached SPM α-$Fe_2O_3$ (Im-IONP) was investigated using TGA-DSC in Argon gas. Figure 5b shows TG, DTG and DSC profile of Im-IONP in the temperature range of 20 °C to 1000 °C. TG curve clearly indicates an initial small weight loss of 4% at the temperature below 175 °C due to desorption of water molecules that might be present on surface of IONP. Higher weight loss of around 11% is observed in the temperature range of around 175 °C to 218 °C. As FTIR analysis clearly shows the absence of adsorption bands around 3400 cm$^{-1}$ corresponding to –OH vibration of water molecules, the weight loss at this temperature range of around 200 °C might be due to thermal degradation of 2 methyl imidazole ligands that surround the IONPs. This thermal degradation of the organic ligands lasts for a wide temperature range from 218 °C to 490 °C, resulting to the weight loss of 12%. It is expected that the decomposition of surfactants is often accompanied with the release of reducing gases such as CO and $H_2$[47]. During the calcination, these gases are believed to play a role in reducing the SPM α-$Fe_2O_3$ NPs, emergence of γ-$Fe_2O_3$ phase and even to the metallic Fe [48]. Further weight loss of 23% is observed in the temperature range of 490 °C to 600 °C.

Another important thermal property is the temperature corresponding to the maximum rate of weight loss ($T_{min}$), which is defined as the peak value of the first derivative of the TG curve. The presence of significant exothermic peak in the DTG profile around 200 °C suggests that the degradation of the imidazole takes place in this temperature range, which is in agreement with weight loss in TG result reported by Ullah *et al.*[49] Therefore, the first stage of decomposition may be from the elements at the tails of the ligands attached to IONP and the second stage decomposition at 560 °C is due to the releasing of the carbon deposited on the surface of iron oxide. This analysis further indicates that carbon is still present in the hetero-IOCC and homo-IOCC as both the samples were calcined in the temperature well below range corresponding to the breakdown of carbon. The presence of carbon is also ensured from the XPS result.

3.4 *Dye removal capacity*

The dye removal percentage and adsorption capacity of the hetero-IOCC in comparison with homo-IOCC and Im-IONPs is shown in Figure 6(a-c). The discoloration of the dye indicates that hetero-IOCC, homo-IOCC, and Im-IONP samples were effective for wastewater treatment. At same experimental condition, the CR dye adsorption capacity for hetero-IOCC was found to be higher (45.84 mg g$^{-1}$) than that of homo-IOCC (5.43 mg g$^{-1}$) and Im-IONP (40.44 mg g$^{-1}$). Impressively, CV dye adsorption capacity of hetero-IOCC was also higher (35.45 mg g$^{-1}$) compared to that for homo-IOCC (9.41 mg g$^{-1}$) and Im-IONP (25.11 mg g$^{-1}$). The adsorption property of CV and CR dyes for the hetero-IOCC is compared with the existing reports and shown in Table 2. The obtained adsorption result certainly elucidated the importance of carbon content presence in the composite over the polymer capping on the NP surface, presence of heterophased grain boundaries over the homophased counterpart and the presence of both α-$Fe_2O_3$ and γ-$Fe_2O_3$ phase over the single phase in the nanocomposite. The creation of well-defined heterophased grain boundaries may lead to creation of more defects mainly positively charge oxygen vacancies, which is in agreement with the XPS result as well. Thus, when heterophased structures are developed, positive charge potential along the grains might be enhanced, thereby inducing a negative charged space layer owing to space charge effect, as shown in **Scheme 2**. So, the positively charge potential along grain boundary core (GB Core) creates the active sites for adsorbing anionic dyes while the neighbouring negatively charged space layer could be the reason for attaching cationic dyes[50]. While bulk properties are dominant in micro-level, the effect of charged state along grain boundaries are more pronounced/significant in nano-level. It has been reported that the grain boundary effect play dominant role in the dye adsorption [51]. The comparatively low removal capacity of Im-IONP sample is due to the fact that α-$Fe_2O_3$ consist of a single grain capped with organic imidazole ligands, thereby reducing the interaction of dye with iron oxide NPs. Hetero-IOCC samples, on the other hand, exhibit well-defined grain boundaries with multiple phases of α-$Fe_2O_3$ & γ-$Fe_2O_3$. The grain boundaries consist of huge defects and the boundaries of different crystallite phase may consist of higher defects compared to the single-phase boundaries. The defects at the boundaries also act as the active sites to adsorb the dye. The surface of Hetero-IOCC have higher exposure and has superior removal capacity than that of Im-IONP. On other hand, Im-IONP and hetero-IOCC sample has larger surface-to-volume ratio compared to homo-IOCC, which is evident from the TEM result of particle size comparison between these samples. Moreover, hetero-IOCC has another advantage of greater saturation magnetization than Im-IONP and homo-IOCC samples that it is sensitive to the external magnetic field and can be separated with ease to use for another cycle.

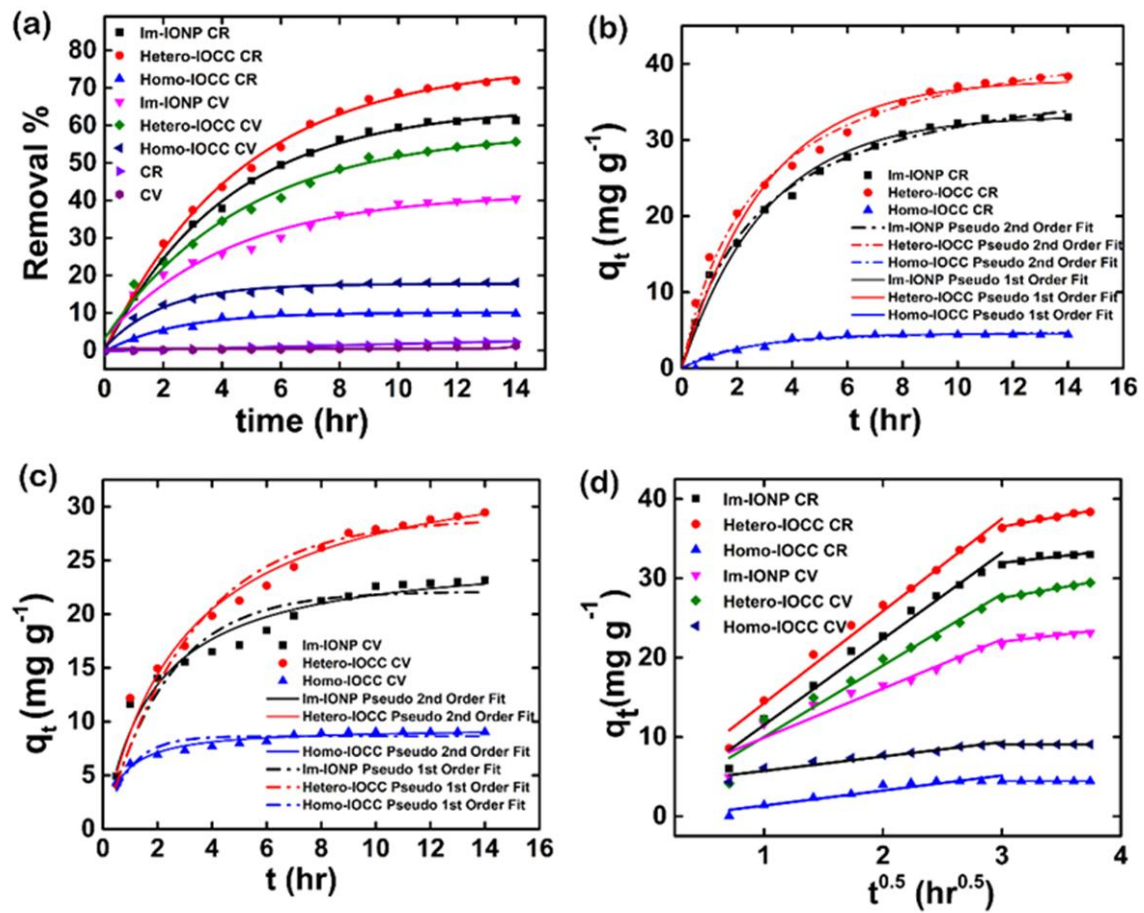

**Figure 6:** (a) Removal percentage of CR and CV dyes for hetero-IOCC and compared with Im-IONP and Homo- IOCC; Pseudo-first-order kinetic plot and Pseudo-second-order kinetic plot of (b) CR (c) CV (d) Weber-Morris plot.

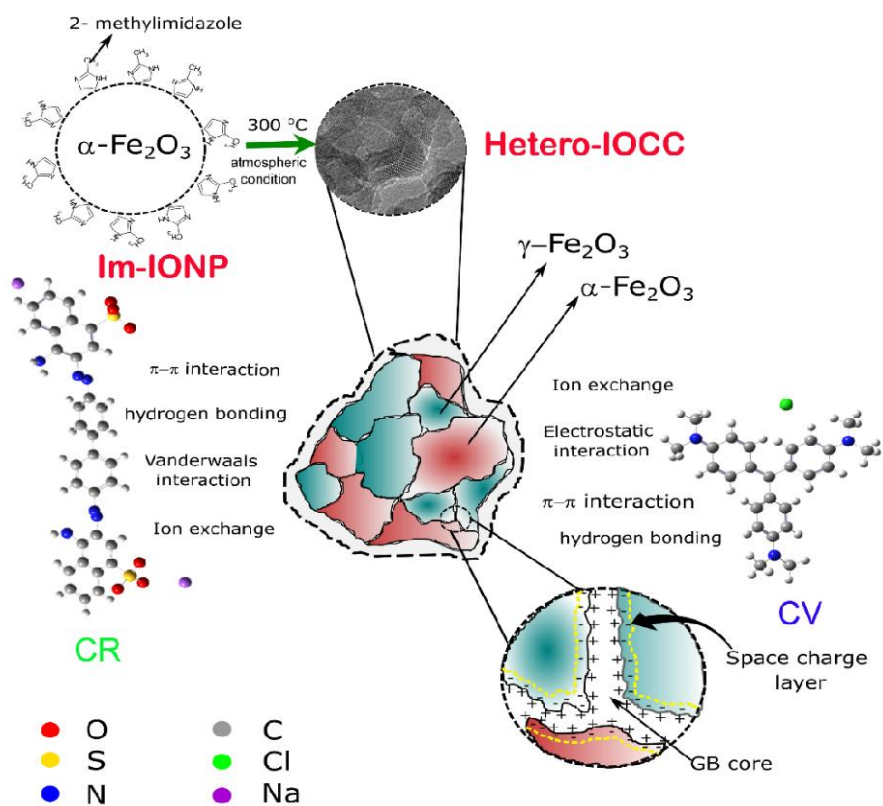

**Scheme 2**: Schematics of proposed dye adsorption mechanism of hetero-IOCC

**Table 2:** Comparison of the maximum adsorption capacities ($q_{max}$) of various reported sorbents for CV and CR. (BTCA: Benzenetetracarboxylic acid, SPION: Super paramagnetic Iron Oxide Nanoparticle. IO: Iron Oxide, PANI: polyaniline, BNNS: Boron nitride nanosheets)

| Sorbent | Synthesis method | Dye | pH | Dye concentration (ppm) | Catalyst dose (g/L) | $q_{max}$ (mg/g) | Removal % | Ref. |
|---|---|---|---|---|---|---|---|---|
| Jute Fibre Carbon | Carbonisation | CV | 8 | 40 | 0.2 | 27.99 | | [52] |
| Non-Magnetic IO | Entrapment method | | 9 | 5 | 0.1 | 16.50 | | [53] |
| IO based on κ-carrageena.n-g poly (methacrylic acid) nano composite | Co-ppt | | 8 | 50 | 0.5 | 28.24 | | [54] |
| BTCA-Fe$_3$O$_4$ | Co-ppt | | 10 | 20 | 0.02 | | 14 | [55] |
| SPION | Co-ppt | | 9 | 10 | 8 | | 20 | [56] |
| **Hetero-IOCC** | **Co-ppt and Calcination** | | **7** | **10** | **0.2** | **35.45** | **53.40** | Present work |
| **Homo-IOCC** | | | | | | 9.41 | 18.60 | |
| **Im-IONP** | **Co-ppt** | | | | | 25.11 | 41.40 | |
| Pretreated Fe$_3$O$_4$ | Co-ppt | CR | 6.6 | 30 | 4 | 21.28 | | [57] |
| Activated carbon | Ultrasound assisted method | | 4 | 200 | 0.2 | 14.92 | | [58] |
| Au-Fe$_3$O$_4$-NCs-AC | | | 4 | 200 | 0.2 | 21.05 | | |
| Activated carbon (ultrasound) | | | 4 | 200 | 0.2 | 24.65 | | |
| BNNS@ Fe$_3$O$_4$ | High Pressu-re | | | 200 | 0.33 | | 39 | [59] |
| PANI/ Fe$_3$O$_4$ | Polymerisation & Mixing | | | 20 | 0.1 | 21.38 | | [60] |
| MnFe$_2$O$_4$ | Co-ppt | | 6.6 | 30 | 4 | 25.78 | | [57] |
| Fe$_3$O$_4$@SiO$_2$ Magnetic NPs | Co-ppt | | 5.5 | 30 | 2 | 24 | | [61] |
| **Hetero-IOCC** | **Co-ppt &** | | 7 | 10 | 0.2 | **45.84** | **74.40** | **Present** |
| **Homo-IOCC** | **Calcination** | | | | | **5.43** | **10.10** | **work** |
| **Im-IONP** | **Co-ppt** | | | | | **40.44** | **60.70** | |

3.4 Adsorption Kinetic study

In order to better understand the adsorption kinetics, the order of the adsorption process was determined by fitting the experimental data into the non-linear pseudo-first-order (eqn 3) and non-linear pseudo-second-order equations (eqn 4):[62]

$$q_t = q_e(1 - e^{-k_1 t}) \quad (3)$$

$$q_t = \frac{k_2 q_e^2 t}{1 + k_2 q_e t} \quad (4)$$

where $q_t$ and $q_e$ are adsorption capacity at time t and at steady state, respectively. $k_1$ and $k_2$ are first order and second order kinetic rate constants, respectively.

To study the diffusion model, the Weber-Morris model equation is used, which is given below:[63]

$$q_t = k_3 t^{1/2} + I \quad (5)$$

where $k_3$ is the rate constant of intra particle diffusion model, $I$ is a constant for any experiment (mg/g).

To analyse the parameters of the pseudo-first-order and pseudo-second-order kinetic models, the non-linear plots of $q_t$ versus time as shown in **Figure 6b** and **6c**. The coefficients of determination ($R^2$), derived from the fitting models have been provided in **Table 3** to show differences in goodness of fitting. In this kinetic study, pseudo-second-order model fitted the kinetics data in the best way (slightly higher $R^2$ value) as compared to the pseudo-first-order kinetic model (with low $R^2$ value), which are shown in **table 3** and **table 4**. This analysis of kinetic data suggests that the pseudo-second-order kinetic model is the better model to depict CV and CR dye adsorption on hetero-IOCC, homo-IOCC and Im-IONP. This indicates that chemisorption could be the dominant factor leading to adsorption of dyes[64].

Table 3 Parameters of Pseudo second order kinetic models for the adsorption of CV and CR dyes onto hetero-IOCC, homo-IOCC and Im-IONP.

| SAMPLE | PSEUDO FIRST ORDER | | | PSEUDO SECOND ORDER | | |
|---|---|---|---|---|---|---|
| | $Q_e$ (mg.g$^{-1}$) | $K_1$ (g.mg$^{-1}$hr$^{-1}$) | $R^2$ | $Q_e$(mg.g$^{-1}$) (calc) | $K_2$ (g.mg$^{-1}$hr$^{-1}$) | $R^2$ |
| Hetero-IOCC CV | 28.94 | 0.31 | 0.90 | 35.45 | 0.01 | 0.98 |
| Homo-IOCC CV | 8.64 | 0.99 | 0.81 | 9.41 | 0.16 | 0.96 |
| Im-IONP CV | 22.11 | 0.42 | 0.89 | 25.11 | 0.02 | 0.96 |

| Hetero-IOCC CR | 37.94 | 0.33 | 0.90 | 45.84 | 0.0083 | 0.93 |
| Homo-IOCC CR | 4.56 | 0.40 | 0.94 | 5.43 | 0.085 | 0.95 |
| Im-IONP CR | 33.21 | 0.32 | 0.94 | 40.44 | 0.0089 | 0.95 |

Table 4: Parameters of Weber-Morris model for the adsorption of CV and CR dyes onto hetero-IOCC and homo-IOCC and Im-IONP

| SAMPLE NAME | STEP-I | | STEP-II | |
|---|---|---|---|---|
| | $K_3$ [mg.(g hr$^{0.5}$)$^{-1}$] | I (mg.g$^{-1}$) | $K_3$ [mg.(g hr$^{0.5}$)$^{-1}$] | I (mg.g$^{-1}$) |
| Hetero-IOCC CV | 9.01 | 1.00 | 19.65 | 0.37 |
| Homo-IOCC CV | 1.81 | 3.91 | 4.42 | 0.01 |
| Im-IONP CV | 6.20 | 3.68 | 16.78 | 1.74 |
| Hetero-IOCC CR | 11.62 | 2.62 | 28.50 | 2.67 |
| Homo-IOCC CR | 1.89 | 0.54 | 9.05 | 0.01 |
| Im-IONP CR | 10.85 | 0.64 | 26.82 | 1.71 |

In the kinetic study of dye removal, the pseudo-first-order and the pseudo-second-order models helped in identifying the adsorption process, however they could not identify the mass transfer and diffusion mechanism. So, Weber-Morris model of intra-particle diffusion model was adopted, as given in eqn 5. Weber-Morris plot of adsorption of CV and CR ($q_t$ vs $t^{0.5}$) is plotted in **Figure 6d** and corresponding fitting parameters are given in **table 4**. Weber-Morris plots of Im-IONP, hetero-IOCC and homo-IOCC samples comprises of 2 steps: (i) external surface adsorption or the diffusion of CV and CR molecules to the surface of adsorbent materials, which continues till the exterior surface reached the saturation and then (ii) the internal diffusion of dye molecules within the pores, also known as intraparticle diffusion, reaching the final equilibrium step [65]. The slope of the linear part of each curve could give the rate constants and the intercepts (I) could be obtained from the extrapolation of the first step in the curves to the time axis (**Figure 6d**). It was found that the extrapolation of initial linear graphs do not pass through the origin, which strongly indicate that the influence of boundary layers on the adsorption process.

3.5 Adsorption mechanism

To analyse the adsorption mechanism of CR and CV dyes, FTIR spectra of hetero-IOCC, homo-IOCC and Im-IONP before and after dye adsorptions were recorded, as shown **in Figure 7**. The peaks located at ~1065 cm$^{-1}$ corresponds to the vibration of sulfonic group of CR. This vibrational band is observed clearly in the FTIR peaks of hetero-IOCC, homo-IOCC and Im-IONP, which are examined after the adsorption of CR dye. Also ~1455 cm$^{-1}$ and ~1585 cm$^{-1}$ vibrational bands which belong to –NH$_2$ and -N=N aromatic group respectively affect the vibration bands of the Im-IONP after adsorption [66]. Moreover, the peak intensities of the CR dye which are observed in the FTIR spectra of the sample after adsorption are reduced. These findings point to the adsorption of CR dye onto the surface of the Im-IONP. Electrostatic interaction between the negative charge sulfonic group and H$^+$ forming SO$_3$H at around 1169 cm$^{-1}$ which is commonly observed in the CR dye adsorption was not observed. Similarly, the vibrational bands at ~1160 cm$^{-1}$, ~1296 cm$^{-1}$, ~1351 cm$^{-1}$, ~1580 cm$^{-1}$, ~2923 cm$^{-1}$ and ~3303

cm$^{-1}$ correspond to stretching vibration of C-H in aromatic ring, C-N vibration, C=C stretching vibrations in aromatic nuclei, N-H stretching vibration of amino group and –OH bond of water adsorbed on CV respectively. These vibrational bands were found to affect the FTIR spectra of hetero-IOCC, homo-IOCC and Im-IONP taken after the adsorption, which confirm the adsorption of CV. The possible mechanism underlying the adsorption process of CV and CR onto hetero-IOCC, homo-IOCC and Im-IONP, as shown in Scheme 2, mainly involve *van der Waals* and hydrogen bonding, which could be related with shifting of the aromatic vibrational bands. Moreover, ion exchange, π-π interactions and hydrophobic interactions may also be involved in the process.

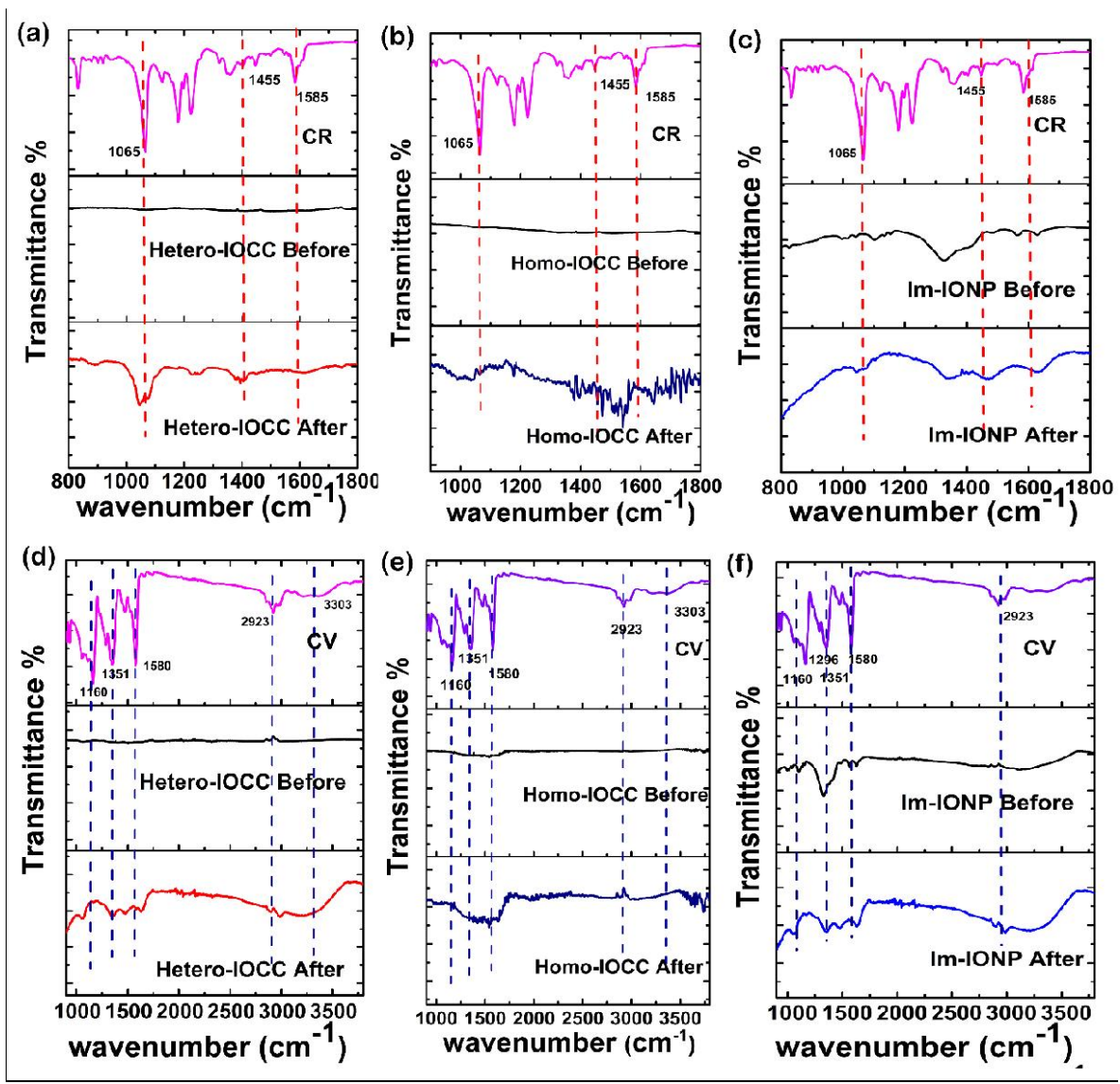

**Figure 7**: FTIR of hetero-IOCC, homo-IOCC and Im-IONP samples before and after adsorption for (a), (b) & (c) CR [in the range of 800 cm$^{-1}$ to 1800 cm$^{-1}$] (c), (d) & (e) CV dyes [in the range of 900 cm$^{-1}$ to 3800 cm$^{-1}$]

**4. Conclusion**

In conclusion, iron-oxide/carbon composite with heterophased grain boundaries (hetero-IOCC) is successfully synthesized by calcining the imidazole-capped superparamagnetic α-Fe$_2$O$_3$ (Im-IONP), which was obtained from the co-precipitation of iron nitrate and 2 methyl imidazole solutions. With existence of α-$Fe_2O_3$ and γ-$Fe_2O_3$ phases, the hetero-IOCC comprises of higher surface-to-volume ratio, less agglomerated nanoparticles (NPs), and greater saturation magnetization compared to the im-IONP and IOCC with homophased grain boundaries (homo-IOCC). The spectroscopic studies elucidates that the oxidation to α-$Fe_2O_3$ favors the agglomeration, whereas the γ-$Fe_2O_3$ were observed to have preserved the size experiencing the superparamagnetic properties. While implemented as an active material for dye adsorption, hetero-IOCC exhibited superior dye adsorption with a capacity of 35.45 mg $g^{-1}$ and 45.84 mg $g^{-1}$ for both cationic Crystal Violet and anionic Congo Red, respectively compared to the in-IONP (25.11 mg $g^{-1}$ and 40.44 mg $g^{-1}$) and homo-IOCC (9.41 mg $g^{-1}$ and 5.43 mg $g^{-1}$). The enhanced dye adsorption in hetero-IOCC is attributed to the (i) heterophase grain boundaries among the α-$Fe_2O_3$ and γ-$Fe_2O_3$ which provide many active sites, (ii) presence of carbon coating which protect NPs from the agglomerations, (iii) the charged defects that got induced on the grain boundaries facilitate both the surface and intraparticle diffusion of both the cationic and anionic dyes. We anticipate that the strategy can be adopted to design and synthesize unique metal oxide/carbon composites with heterophased grain boundaries for a wide range of applications such as catalysis, gas sensors, hydrogen evolution reactions, energy storage and sustainable wastewater treatment.

**CRediT authorship contribution statement**

**K Priyananda Singh:** Conceptualization, Data curation, Formal analysis, Writing original draft, review & Editing; **Boris Wareppam:** Investigation, review & Editing; **Raghavendra K G:** Investigation; **N. Joseph Singh:** Formal Analysis; **A. C. de Oliveira:** Investigation; **V. K. Garg:** Investigation; **S.Ghosh:** Investigation, supervision, review and editing; **L. Herojit Singh:** Supervision, Methodology, Investigation, Project administration, Writing and review; Resources, Visualization, Funding acquisition.

**Declaration of Competing Interest**

The authors declare that they have no known competing financial interests or personal relationships that could have appeared to influence the work reported in this paper.

**Acknowledgement**

K. P. S. would like to extend gratitude to UGC for providing the fellowship and financial support. The author LHS would like to thank DST-SERB for the financial assistance under the project having file no. CRG/2021/001611. S.G. acknowledge Marie Skłodowska–Curie postdoctoral fellowship (101067998 — ENHANCER). The authors thank the Department of Chemistry, NIT Manipur for the XRD and FTIR measurements. The author also thanks Daniel Cliff Gonmei, Department of Physics, NIT Manipur for his valuable support in the synthesis process. The author also would like to acknowledge SAIF, IIT Bombay for providing HRTEM facility for my research work. The advanced characterization facilities – VSM and TGA/DTA provided by Centre for Instrumental Facility (CIF) at IIT Guwahati and XPS provided by Institute Instrumentation Centre at IIT Roorkee are also acknowledged.